\documentclass[12pt]{article}
\usepackage{amssymb}
\usepackage{amsmath, amsthm}
\newcommand{\be}{\begin{equation}}
\newcommand{\ee}{\end{equation}}
\begin{document}
\begin{titlepage}
\title{No ``big trips'' for the universe}
\author{Valerio Faraoni\\ \\
{\small \it Physics Department, Bishop's University}\\
{\small \it 2600 College St., Sherbrooke, Qu\`{e}bec, Canada J1M~0C8}
}
\date{} \maketitle
\thispagestyle{empty}
\vspace*{1truecm}
\begin{abstract}
It has been claimed in several papers that a phantom 
energy-dominated universe can undergo a ``big trip'', i.e.,  
tunneling through a wormhole that grows faster than the cosmic substratum due to the accretion of phantom energy, and 
will reappear on the other mouth of the wormhole. We show that  
such claims are unfounded and contradict the
Einstein equations.  
\end{abstract} \vspace*{1truecm}
%\noindent
\begin{center} {\bf PACS:} 04.20.-q, 04.20.Jb, 98.80.-k   
\end{center}
\begin{center}  {\bf Keywords:} wormholes, phantom energy, Big Rip.
\end{center}
\setcounter{page}{1}
\end{titlepage}

%\def\theequation{\arabic{section}.\arabic{equation}}

%%%%%%%%%%%%%%%%%%%%%%%%%%%%%%%%%%%%%%%%%%%%%%%%%%%%%%%%%%%%%%%%%%%%%%%%%%%%%%%%%%%%

%\section{Introduction}
\setcounter{equation}{0}
\setcounter{page}{2}
The modern picture of the accelerating universe has led to many 
dark energy models 
accounting for the observed acceleration. While observational support from  
distant supernovae of type Ia \cite{SN} is still marginal, models incorporating a 
phantom energy component 
(i.e., a form
of energy with pressure $P<-\rho$, where $\rho$ is the energy density) have received the 
attention of many authors \cite{phantom}. Such models may exhibit a Big Rip, a singularity 
occurring at a 
finite time in the future at which the scale factor, the energy density, and the pressure 
diverge \cite{BigRip}.  In the context of such models it has recently been proposed that  the 
universe could avoid the Big Rip  by tunneling though a wormhole contained in it, 
disappearing through one mouth and reappearing from the other mouth 
\cite{PRL}, a process 
dubbed ``big trip''. The mechanism by which the big trip would be achieved would be the 
catastrophic growth of a wormhole by accretion of phantom energy at a speed larger than the 
expansion rate of the cosmic substratum. In 
Ref.~\cite{FaraoniIsrael} it was shown that  the original claim of such a possibility is 
unfounded, not being based on actual calculations, and explicit exact solutions were 
presented describing 
wormholes embedded in a Friedmann-Lemaitre-Robertson-Walker (hereafter FLRW) universe and 
accreting phantom energy from the cosmic fluid. While the universe approaches the Big Rip, 
these rather general classes of wormhole solutions end up expanding and becoming comoving, even 
if they initially expand with arbitrary velocity relative to the cosmic fluid 
\cite{FaraoniIsrael}. Another solution of this kind was found 
in Ref.~\cite{SushkovKim}. 
However, more recent claims that the big trip is possible have appeared in the literature 
(\cite{somenotes,wormholeaccretion} --- see also 
\cite{phantominflation}-\cite{Jimenez}); 
these are again not 
substantiated by calculations and reflect more the wishful thinking of these authors 
than real results. Here we want to settle the issue of the big trip.

In his  recent claim of a big trip  \cite{somenotes}, as well as 
in the first 
paper containing such a claim  \cite{PRL}, Gonzalez-Diaz uses 
spherically symmetric 
{\em static} metrics to deduce the possibility of the big trip. 
Ref.~\cite{somenotes} is 
structured as a  reply to potential or published objections to the possibility of the big 
trip. Aside from the difficulty for stable macroscopic wormholes to exist, the first and in 
our view, the most serious, objection \cite{FaraoniIsrael} is 
that a static 
metric can not possibly describe an extremely rapid, 
catastrophic accretion process. 
Ref.~\cite{somenotes} is flawed in this regard, as is shown in the following. This 
paper begins with the static Morris-Thorne wormhole metric  
\cite{MorrisThorne} with zero shift 
function (i.e., $g_{00}=-1$)
\be\label{1}
ds^2=-dt^2+\frac{1}{1-\frac{K(r)}{r}} dr^2+r^2d\Omega_2^2 \;,
\ee
where $d\Omega_2^2$ is the metric on the unit two-sphere. 
The Einstein equations for a spherically symmetric metric 
of the form 
\be \label{sphmetric}
ds^2=-e^{\nu(t,r)} dt^2+ e^{\lambda(t,r)} dr^2 +r^2 d\Omega_2^2
\ee
assume the form \cite{LandauLifschitz}
\begin{eqnarray}
&& \mbox{e}^{-\lambda}\left( \frac{\nu'}{r}+\frac{1}{r^2} 
\right) -\frac{1}{r^2}=8\pi G  T^1_1 \;, \label{3}\\
&&\nonumber \\
&& \frac{\mbox{e}^{-\lambda}}{2} \left( \nu''+\frac{\left( \nu ' 
\right)^2}{2} +\frac{\nu'-\lambda'}{r} -\frac{\nu'\lambda'}{2} 
\right) -\frac{e^{-\nu}}{2}  \left( \ddot{\lambda}   +\frac{ 
\left( \dot{\lambda} \right)^2}{2}  
-\frac{\dot{\lambda}\dot{\nu}}{2}   \right)=8\pi G \, T_2^2 
\nonumber \\
&& = 8\pi G \, T_3^3 \;, \label{4}\\
&&\nonumber \\
&& \mbox{e}^{-\lambda} \left( \frac{1}{r^2} - \frac{\lambda 
'}{r}  \right)  -\frac{1}{r^2}=8\pi G \, T_0^0 \;, 
\label{4bis}\\
&&\nonumber \\
&& \mbox{e}^{-\lambda} \, \dot{\lambda}=8\pi G r \, T_0^1 \;, 
\label{5}
\end{eqnarray}
where $T_{ab}$ is the energy-momentum tensor, $\dot{}\equiv \partial /\partial t$, and $'\equiv 
\partial/\partial r$.  For the metric (\ref{1}), eq.~(\ref{5})  
is equivalent to
\be\label{6}
\dot{K}= 8\pi G \left( r-K \right)^2T_0^1 \;.
\ee
Therefore, a static metric which corresponds to $\dot{K}=0$ 
automatically implies $T_0^1=0$ and vice-versa. There is no 
radial energy flux 
and no accretion of phantom energy onto the wormhole, let alone accretion of the entire 
universe!  The use of equations derived from a  static metric can not be justified for extreme 
accretion regimes. Historically, this point was not missed by McVittie when he introduced 
in his 1933 paper \cite{McVittie} the 
well known McVittie solution  
\be\label{7}
ds^2=-\left( \frac{ 1-\frac{m(t)}{2r}}{1+\frac{m(t)}{2r} } \right)^2 dt^2+ a^2(t) \left( 
1+\frac{m(t)}{2r} \right)^4 \left( dr^2 + r^2d \Omega_2^2 \right)
\ee
describing a spherically symmetric central object embedded in a FLRW cosmological background 
in isotropic coordinates (here we limit ourselves to the spatially flat case relevant in the 
context of 
Refs.~\cite{somenotes, PRL, wormholeaccretion, phantominflation, 
GonzalezSiguenza, Jimenez}). The central mass satisfies 
\be\label{mdot}
\frac{\dot{m}}{m}=-\, \frac{\dot{a}}{a} \;,
\ee
which follows from the equation ${T_0}^1=0$, which in turn is derived from the explicit 
assumption that the central object does not accrete the cosmic fluid (assumption {\em e)} of
Ref.~\cite{McVittie}). This solution of the Einstein equations 
shows that, in fact, in order to 
have accretion not only the metric must be time-dependent, but also the energy-momentum 
tensor must have a mixed component ${T_0}^1$ describing the radial energy flow onto the 
central object, and missing in a static metric such as 
(\ref{1}). Then, 
eq.~(\ref{mdot})  will contain an  extra term describing this 
radial flow. While no change in the 
mass of the 
wormhole, and therefore a static metric, could be reasonable 
approximations for slow  accretion rates, Gonzalez-Diaz claims 
that their use is justified in the extreme accretion 
regime that he wants to describe, which contradicts the Einstein 
equation~(\ref{5}).   He states that ``... 
even though the starting metric is static, the energy stored in the wormhole  must change 
with time by virtue of dark energy accretion ...'' 
\cite{somenotes}: it is impossible to have $ T_0^0, T_1^1, 
T_2^2$, and $T_3^3$ time-dependent on the right hand sides of 
eqs.~(\ref{3})-(\ref{4bis})  
and simultaneously have the left hand sides (containing $\lambda$ and its derivatives)  
time-independent.  
Refs.~\cite{somenotes,wormholeaccretion,PRL,GonzalezSiguenza} do not 
propose  a single solution of the Einstein equations but rather try to deduce the growth rate 
of the wormhole mass by modifying formulas valid for static 
solutions in an inconsistent way, as shown in the following. 

The discussion of \cite{somenotes}  is based on the covariant 
conservation equation $\nabla^b T_{ab}=0$ taken from 
Ref.~\cite{BDE} (hereafter ``BDE'') and replacing the final 
result of this reference 
containing a $K(r)$ with a time-dependent $K(t,r)$. Now, the 
equations of \cite{BDE} were derived for:\\
1) a $T_{ab}$ describing a non-gravitating {\em test fluid}, 
which can not represent the cosmic fluid causing the expansion 
of the universe and the Big Rip;\\
2) the static {\em Schwarzschild metric} corresponding to {\em 
vacuum} (consistent with assumption 1)).
One needs instead a metric describing simultaneously the 
central object (wormhole) {\em and} the cosmological background 
universe in which it is embedded, \'a la McVittie. The 
conservation equation $\nabla^b T_{ab}=0$ used  in 
\cite{somenotes} refers to the Schwarzschild metric describing 
only a central black hole. While the last metric presented in 
\cite{somenotes} could in principle achieve the goal, no 
calculations 
pertaining to it are presented.  Now, the first conservation 
equation of \cite{somenotes} corresponds to eq.~(3) of 
\cite{BDE} and can only be 
obtained in  {\em vacuum} (i.e., with $T_{ab}$ describing a 
non-gravitating  test fluid that is accreted) and the 
Schwarzschild metric. In fact, using the general spherically 
symmetric metric (\ref{sphmetric}) the conservation equation 
 projected on the $u^a$ direction, $u^a \nabla_b {T_{a}}^b=0$ 
becomes
$$
\frac{\rho'}{P+\rho} +\frac{u^t}{u^r} \, 
\frac{\dot{\rho}}{P+\rho}+\frac{u^t}{u^r}\left( 
\frac{\dot{\nu}+\dot{\lambda}}{2}\right) +\frac{  \partial_t 
u^t}{u^r}+\frac{\lambda'+\nu'}{2} +\frac{2}{r}+\partial_r 
\left( \ln u^r\right)=0 \;.
$$
This can  be simplified to eq.~(3) of BDE used in  
\cite{somenotes}  only 
under the assumptions of vacuum, $\partial_t u^a=0$, 
$\dot{\rho}=0$, and the metric is then forced to be  
Schwarzschild. Then, $\lambda=-\nu$  \cite{LandauLifschitz} and 
we obtain
$$
r^2 u^r \, e^{\int \frac{d\rho}{P+\rho}}=C 
$$
(eq.~(3) of BDE). 
Ref.~\cite{somenotes}  neglects the derivatives  
$\dot{\nu},\dot{\lambda},  \nu',\lambda', \dot{\rho}$  and $u^t$ 
and simply replaces  $K(r)$ with $K(t,r)$ in the result valid 
for a vacuum static metric, and the metric  proposed does not 
solve the Einstein equations.

Second, the conservation equation  $\nabla^bT_{ab}=0$ alone is 
certainly not sufficient to determine the metric; it determines 
the dynamics of (test or gravitating) matter if the metric is 
given.  This is exactly the point of view of the 
authors of Ref.~\cite{BDE}, who 
assume a  Schwarzschild metric and compute 
the accretion rate of a 
{\em test fluid on the Schwarzschild black hole}. Gonzalez-Diaz 
takes the conservation equation from \cite{BDE} and tries to 
determine the  metric from $\nabla^b T_{ab}=0$ alone, which is 
impossible. In 
reality the metric is still Schwarzschild but 
in \cite{somenotes} it is proposed to  alter it {\em a 
posteriori} by adding time dependence 
in $ K(r)$. However this incorrect procedure neglects most of 
the relevant terms in  the Einstein equations.

A general proof of the non-existence of the big trip is, of 
course, impossible, because one does not know the most general 
solution of the Einstein equations representing a wormhole, even 
for spherical symmetry. This is due to the failure of Birkhoff's 
theorem in non-vacuum, non-asymptotically flat spacetimes and to 
the lack of a precise definition of wormhole\footnote{Indeed, 
very few exact solutions describing spherical objects (even 
black 
holes) embedded in a cosmological background are known, 
including 
the McVittie solution (\ref{7}) which is singular at $r=m/2$ 
except when it reduces to the Schwarszchild-de Sitter black hole 
\cite{Nolan}.}. We can, however, restrict to metrics of 
the form  (\ref{1}) with $K$ promoted to a function 
of {\em both} $t$ and $r$, which is considered in 
\cite{somenotes}. Then, $\nu\equiv 0 $ and 
$\mbox{e}^{\lambda} =\left( 1-\frac{K}{r} \right)^{-1}$. Let us 
consider a perfect fluid with 
stress-energy tensor $T_{ab}=\left( P+\rho \right) u_a u_b 
+Pg_{ab} 
$, where $u^a=\left( u^0, u, 0,0 \right)$ is the fluid 
four-velocity wit radial component $u\equiv u^r<0$ describing  
inflow onto the wormhole. The normalization $u^c u_c=-1$ yields 
$ u^0= \mbox{e}^{\lambda} \, \left|u \right| $. The {\em proper} 
velocity of the fluid is 
\be
v\equiv \frac{ dr_{phys}}{d\tau}=\left( 1-\frac{K}{r} 
\right)^{-1}  \frac{dr}{d\tau}=\frac{u}{1-\frac{K}{r} } \;,
\ee
where $dr_{phys}=\sqrt{g_{rr}}\, dr $ is the proper distance 
in the radial direction and $\tau=t$ is the proper time. 
By combining eqs.~(\ref{3}) and (\ref{5}) and the expression of 
$v$ one obtains
\be\label{mixa}
\frac{\lambda ' \mbox{e}^{\lambda} }{ \mbox{e}^{2\lambda}-1} 
=8\pi 
G r \left| P+\rho \right| v^2 \;.
\ee 
After straightforward manipulations of the left hand side, 
eq.~(\ref{mixa}) is rewritten as 
\be 
\frac{d}{dr} \left[ \ln \left( \frac{ 
\mbox{e}^{\lambda} +1 }{ \left| \mbox{e}^{\lambda}-1 \right| 
} \right)\right] 
=16\pi G r \left| P+\rho \right| v^2 \;.
\ee
Assuming, as done in \cite{somenotes}, that 
$\left(P+\rho\right)=\left( w+1 \right) \rho=\left( w+1 \right) 
\rho_0 a^{-3(w+1)}$ in a universe dominated by phantom energy 
with equation of state $P=w\rho$ ($w<-1$) heading toward the Big 
Rip, the right hand side diverges unless $v \sim 
a^{\frac{3(w+1)}{2}} $, 
which implies that 
\be
\ln \left( \frac{ 
\mbox{e}^{\lambda}+1}{\left| \mbox{e}^{\lambda}-1 \right| } 
\right) =16\pi G \int dr \, r \left| P+\rho \right| v^2 
\ee
diverges as the Big Rip is approached, or that $K(t,r) 
\rightarrow 0 $ for all values of $r$  in this limit; but then 
the function $K$ 
disappears from the metric (\ref{1}), which becomes the 
Minkowski metric! This conundrum is avoided only if $v\sim 
a^{\frac{3(w+1)}{2} }$,  which guarantees that the right hand 
side of 
eq.~(\ref{mixa}) 
stays finite. But the vanishing of the proper radial velocity of 
the fluid  means that accretion stops asymptotically near the 
Big Rip, and the big trip is again prevented.

In addition to these quantitative arguments and to the explicit 
solutions constructed in  Refs.~\cite{FaraoniIsrael,SushkovKim} 
we remark that, from the purely conceptual point 
of view,  it is hard to give  a meaning to the idea of an  entire 
universe disappearing into one of its parts contained within it. A vague argument 
invoking a multiverse, sketched in Ref.~\cite{somenotes}, fails to address this problem,  
as the author of \cite{somenotes} himself seems to be aware of.

To summarize: here we show that the proposed metrics with the 
big trip property are not solutions of the Einstein equations.  
To study the possibility of a big trip one must examine  
metrics describing  simultaneously 
the wormhole and the universe in which it is embedded (with 
$T_{ab} $ a gravitating fluid) and solve the Einstein 
equations  consistently. This is what is done  in  
Refs.~\cite{FaraoniIsrael,SushkovKim}. In spite 
of the fact that rather large classes of solutions were found,  
none of them has the big trip 
property.  We do not claim that these are {\em all} the possible 
solutions for a wormhole embedded in a phantom energy universe: 
they are just all the solutions known to date. 
Ref.~\cite{somenotes} and the companion papers propose  
guesses on how the equations and the solutions for 
a big trip would look like, and these guesses are incorrect.  It 
has not been shown to date that the big  trip of the universe is 
possible or has physical meaning.

%%%%%%%%%%%%%%%%%%%%%%%%%%%%%%%%%%%%%%%%%%%%%%%%%%%%%%%%%
\section*{Acknowledgment}

This work was supported by the Natural Sciences and Engineering 
Research Council of Canada  (NSERC).

%\clearpage
\vskip1truecm

%%%%%%%%%%%%%%%%%%%%%%%%%%%%%%%%%%%%%%%%%%%%%%%%%%%%%%%%%%%%%%%%%%%%%%%%

\end{document}